\documentclass[
    ,final            
  ]
  {aipproc}

\layoutstyle{6x9}

\begin{document}

\def\ga{\lower.5ex\hbox{$\buildrel>\over\sim$}}
\def\la{\lower.5ex\hbox{$\buildrel<\over\sim$}}

\title{Soft-excess in ULX spectra:\\the chilled-disk scenario}


\author{R. Soria}{
  address={Harvard-Smithsonian Center for Astrophysics, \
  60 Garden st, Cambridge, MA 092138, USA}
}

\author{A. C. Gon\c{c}alves}{
  address={LUTH, Observatoire de Paris-Meudon, 5 Place Jules Janssen, 
92195 Meudon, France}
,altaddress={CAAUL, Observatorio Astronomico de Lisboa, Tapada da
Ajuda, 1349-018 Lisboa, Portugal}
}

\author{Z. Kuncic}{
  address={School of Physics, University of Sydney, NSW 2006, Australia}
}

\begin{abstract}
Soft X-ray spectra of ULXs show small deviations from 
a power-law model, that can be attributed to reprocessing 
in a fast, ionized outflow, or to thermal emission from a cool disk. 
If it is thermal emission, the cool peak temperature can be 
explained by an inner disk that radiates only a small fraction 
of the gravitational power, transferring the rest to an upscattering 
medium which is then responsible for the dominant power-law
component. This scenario does not require intermediate-mass 
black holes: we use a phenomenological model to show 
that the observed X-ray luminosities and spectra 
of ULXs are consistent with typical masses $\sim 50$--$100 M_{\odot}$.
\end{abstract}

\maketitle


\section{Black hole masses from the disk emission}

Determining the masses of the accreting black holes (BHs) 
in ultraluminous X-ray sources (ULXs) is a key 
unsolved problem in X-ray astrophysics. In the absence 
of direct kinematic measurements of their mass functions, 
indirect methods based on X-ray spectral modelling have 
sometimes been used, by analogy with Galactic BH X-ray 
binaries, whose spectra can be approximated, 
in the ``canonical'' $0.3$--$10$ keV range, 
with a thermal component 
plus a power-law. The thermal component is consistent 
with disk-blackbody emission from an optically 
thick disk, while the power-law is generally attributed 
to inverse-Compton scattering of softer disk photons off 
high-energy electrons. The peak temperature and luminosity 
of the thermal component are good indicators 
of the size of the X-ray-emitting inner-disk region, 
which in turn is related to the BH mass. 

In the simplest and most commonly used disk-blackbody 
approximation \cite{mak86},
\begin{eqnarray}
L_0 &\approx& 4 \pi \sigma T_0^4 R_{0}^2 \\
L_0 &=& \eta \dot{M} c^2 \\
\sigma T_0^4 &\approx& \frac{3GM\dot{M}}{8 \pi R_{0}^3},
\end{eqnarray}
where $R_{0}$ is a characteristic size $\approx R_{\rm in}$ 
(inner-disk radius) and $T_0$ is a characteristic temperature 
$\approx T_{\rm max}$, and where we have ignored a factor 
related to the no-torque condition at $R_{\rm in}$ and a hardening
factor. (Those terms can be taken into account later as a numerical
factor $g$ \cite{fab04}.)
$L_0$ is the integrated disk luminosity, under the assumption 
that the disk dissipates most of the gravitational energy 
of the inflow. The radiative efficiency  
$\eta \approx 0.1$--$0.3$ for a source in a high/soft state 
(dominated by a bright disk).

From (1), (2), and (3):  
\begin{equation}
M \approx \frac{c^2 g^2 \eta L_0^{1/2} T_0^{-2}}{3G(\sigma \pi)^{1/2}}\,
 \approx 10.2 \, \left(\frac{g}{1.35}\right)^2 \, \left(\frac{\eta}{0.2}\right)\,
    \left(\frac{L_0}{5 \times 10^{38} 
    {\rm{~erg~s}}^{-1}}\right)^{1/2}\, \left(\frac{T_0}{1  
    {\rm{~keV}}}\right)^{-2} \, M_{\odot},
\end{equation}
which also implies $L_0 \sim M^2 T_0^{4}$. The effective 
hardening factor $g \approx 1.35$ \cite{fab04}. The 
observable quantities $L_0$ and $T_0$ may come from 
X-ray spectral fitting, and the efficiency is known  
within a factor of $2$ based on standard accretion models.
Despite the various approximations, (4) works well 
(within a factor of $2$) when applied to the masses of Galactic BHs. 
It is also observationally verified that the X-ray luminosity 
of each source varies over months or years 
following the relation $L_{\rm X} \sim T_0^{4}$ at constant $M$ 
\cite{mil04}, at least until it approaches the Eddington limit, 
$L_{\rm Edd} \approx 1.3 \times 10^{38} (M/M_{\odot})$ erg s$^{-1}$.

\section{Soft X-ray emission in ULXs} 

ULX X-ray spectra can also be modelled 
with a power-law plus a thermal component, much cooler 
than that observed from Galactic BHs --- typically, with 
temperatures $kT_0 \sim 0.12$--$0.20$ keV for the brightest 
sources. If (4) is applied, with total 
$0.3$--$10$ keV luminosities $\approx 1$--$2 \times 10^{40}$ 
erg s$^{-1}$, the inferred masses are $\sim 1000 M_{\odot}$. 
This argument has been used in support 
of the intermediate-mass BH interpretation of ULXs 
\cite{mil04}; however, it is based on questionable assumptions. 

Firstly, there is little direct evidence 
that the ``soft excess'' in ULXs is due to disk emission, 
partly because of our limited observing band. Analogous 
soft-excesses in AGN, and particularly in narrow-line 
Seyfert 1 galaxies (NLS1) can be explained as blurred, 
ionized absorption (mostly in the $\sim 0.5$--$2$ keV 
band) and reprocessing of the primary 
power-law-like spectrum in a fast outflow \cite{gie04}\cite{che06}. 
We showed \cite{gon06} that a similar interpretation 
can be applied to ULXs, whose X-ray spectra are very 
similar to those of NLS1. We do not speculate here 
why the primary spectrum is non-thermal. As a possible 
analogy, the power-law component becomes dominant over 
the disk component in Galactic BHs in the very high state, 
at luminosities $\ga L_{\rm Edd}$.

Secondly, (4) holds only if the disk is emitting 
most of the energy liberated by accretion, as in (2).
In ULXs, this is not the case. Even if the deviation 
from the power-law spectrum at soft energies is indeed from 
disk emission, the X-ray spectrum is still dominated 
by the power-law component, and the thermal component 
would represent only $\sim 10\%$ of the $0.3$--$10$ keV emission
\cite{sto06}. Thus, (2) and (3) no longer apply in this form.
Physically, this suggests that most of the accretion power 
is not radiated by the disk, but is efficiently transferred 
in other forms (mechanical, thermal or magnetic energy) 
to an upscattering medium, and then partly radiated 
with a non-thermal (power-law-like) spectrum.

In summary, we have {\it two alternative scenarios to describe 
the first-order deviations from a power-law spectrum} 
in the soft X-ray band: either outflow reprocessing (mostly 
ionized absorption), or a modified disk that is 
only radiating a small fraction of the energy released 
by accretion; or a combination of both. 
Both are consistent with the X-ray 
observations. We discuss the ionized outflow  
scenario elsewhere in these Proceedings \cite{gon06b}; 
see also \cite{gon06}. 
Here we outline the main implications of the disk hypothesis.

\section{The chilled disk scenario}

The key physical question for such a scenario is why 
the disk is so cold for its luminosity. Even when we take 
into account only the luminosity in the fitted thermal 
component, ULX disks would radiate up to $\approx$ a few $10^{39}$ 
erg s$^{-1}$ with a much lower peak temperature 
than that observed from Galactic BHs in their high state.
A standard, truncated disk, replaced in the inner region
by a radiatively-inefficient flow, would produce a power-law 
dominated spectrum with a cold disk component, as is the case  
for stellar-mass BHs in the low/hard state. However, if ULXs 
are in the low/hard state at luminosities $> 10^{40}$ 
erg s$^{-1}$, their BH masses should be $\ga 10^4 M_{\odot}$. 
The formation of such massive remnants is difficult to explain 
with existing models of stellar or star cluster evolution.

Therefore, we consider an alternative scenario based 
on the following phenomenological arguments: 
\begin{enumerate}
\item the mass inflow rate
is sufficiently high that the disk extends all the way to the innermost 
stable circular orbit; 
\item hence, the conversion 
of gravitational energy to accretion power is efficient: 
the total power is $\ga 0.1 \dot{M}c^2$; 
\item the outer region, at radii $R \ge R_{\rm c}$, is a standard 
optically-thick disk ($T \sim R^{-3/4}$) 
radiating its accretion power with a multicolour diskbb spectrum;
\item in the inner region, at $R_{\rm in} \le R \le R_{\rm c}$,  
only a small fraction of the released power is directly radiated 
by the disk. The rest is efficiently transferred to a corona 
or jet or magnetized outflow, acting as an upscattering medium.
\end{enumerate}

One possible way to model this process is by assuming 
that a constant fraction of power is extracted from the disk 
at all radii. The alternative possibility is that the outer disk 
region, at radii $R \ge R_{\rm c}$, is a standard 
optically-thick disk ($T \sim R^{-3/4}$), 
while the non-thermal component (jet, magnetized outflow 
or corona) becomes dominant in the inner region, 
at $R_{\rm in} \le R \le R_{\rm c}$.
In any case, the inner disk is {\it cooler 
than a standard disk}, because it radiates only a flux 
\begin{equation}
\sigma T_{\rm eff}(R)^4 \approx \alpha \, \frac{3GM\dot{M}}{8 \pi R^3} - F_{\rm nt}(R) 
   < \frac{3GM\dot{M}}{8 \pi R^3},
\end{equation}
where $(1-\alpha) > 0$ is the constant fraction of power 
extracted from the disk at all radii, 
and $F_{\rm nt}(R)$ is the additional flux removed 
from the innermost region, inside the transition radius 
$R_{\rm c}$ \citep{SoK06}.

If/where $F_{\rm nt}(R) =0$, $T_{\rm eff} \sim R^{-3/4}$, 
and the emitted spectrum is still a disk blackbody. 
If $F_{\rm nt}(R) = 0$ at all radii, the mass estimate (4) holds true, 
independent of $\alpha$. As the accretion rate is varied, 
the disk follows the same $L_0 \sim M^2 T_0^{4}$ track, 
athough cooler and less luminous than in the $\alpha = 1$ case, 
for a given accretion rate.
However, we argue that
the extraction of power from the disk is more likely 
to occur in the inner region, while the disk may be undisturbed 
at large radii.
It follows from (5) that the temperature profile 
flattens out in the inner region, where $F_{\rm nt}$ dominates.
In particular, if $T(R)$ increases more slowly that $R^{-1/2}$ for 
$R \rightarrow R_{\rm in}$, the maximum contribution to the disk emission 
occurs at $R \approx R_{\rm c}$, $T \approx T(R_{\rm c})$. 
Those values will also be proportional (accounting 
for the hardening factor) to the fitted coulor 
peak temperature and radii derived from an observed spectrum. 
For simplicity, here we assume that 
$T = T(R_{\rm c}) =$ constant 
inside $R_{\rm c}$; our conclusions do not depend on the exact 
choice of a temperature law. Moreover, only a fraction 
$\beta < 1$ of the disk emission from the inner region 
may be directly visible; if all the disk photons from that region 
are comptonized, then $\beta \approx 0$. 

We can now re-write (1), (2), (3) as:
\begin{eqnarray}
L_0 &\approx& 4\pi R_{\rm c}^2 \sigma T_{\rm eff}(R_{\rm c})^4 
   + 2\pi \beta (R_{\rm c}^2 - R_{\rm in}^2)\sigma T_{\rm eff}(R_{\rm c})^4 
   \approx 4\pi R_{0}^2 \sigma T_0^4\\
L_0 &=& \alpha f \eta \dot{M} c^2 \\
\sigma T_{\rm eff}(R_{\rm c})^4 &\approx &
     \alpha \frac{3GM\dot{M}}{8\pi R_{\rm c}^3}. 
\end{eqnarray}
For simplicity, we have assumed here that $\beta \ll 1$,   
so that there is effectively no observational 
difference between a disk truncated at $R_{\rm c}$ 
and replaced by an efficient comptonizing medium 
or outflow inside that radius, 
and a disk extending to the innermost stable orbit but 
with a flat temperature distribution. Physically, there is 
of course a significant difference, because the mechanisms 
responsible for extracting accretion power and transferring it 
to a jet or a corona may require the presence of an accretion 
disk in the inner region, even though we cannot see it directly. 
See for example the model of \cite{kun04}, 
where power is removed 
from the disk by the vertical component of the magnetic torque.
In (7), we have defined $\alpha f$ as the fraction 
of accretion power radiated by the disk, 
with $0 < \alpha \la 1$ and $0 < f \la 1$. 
A fraction $(1-\alpha)$ of the accretion power is extracted 
uniformly over the whole disk surface, and a further fraction
$\alpha (1-f)$ is extracted from the disk inside the transition 
radius, as discussed above.
It is difficult to determine $f$ (or, equivalently, 
estimate the flattening of the temperature profile 
in the inner disk) observationally. We can 
estimate $f \sim 0.1$, based on the fitted ratio of soft thermal 
emission over total X-ray luminosity in ULXs. 
If anything, that ratio is an overestimate of $f$, 
because non-thermal processes are less radiatively efficient than
thermal disk emission, and some power is still likely 
to escape in non-radiative forms.
The other two observable parameters, as before, 
are the peak colour temperature $T_0$ and the luminosity $L_0$ 
of the thermal component.

We can now solve (6), (7), (8) for $M$, $\dot{M}$ and $R_{\rm c}$ 
as a function of the parameters $f$, $\alpha$, $g$, 
and of the observable quantities $T_0$ and $L_0$ (or $R_0$): 
\begin{eqnarray}
M &\approx&  90.7  \left(\frac{\eta}{0.2}\right) 
 \left(\frac{f}{0.1}\right) \left(\frac{g}{1.35}\right)^2
 \left(\frac{L_0}{2 \times 10^{39} 
  {\rm{~erg~s}}^{-1}}\right)^{1/2} \nonumber\\
 &\times& \left(\frac{T_0}{0.15~{\rm keV}}\right)^{-2} \, M_{\odot} \\
\dot{M} &\approx& 1.8 \times 10^{-6} \, \left(\frac{0.2}{\eta}\right) 
 \left(\frac{0.1}{f}\right)\, \left(\frac{1}{\alpha}\right)\,
 \left(\frac{L_0}{2 \times 10^{39} 
  {\rm{~erg~s}}^{-1}}\right) \ \ M_{\odot} \ {\rm yr}^{-1}\\ 
R_{\rm c} &\approx& g^2 R_0
  \approx \frac{3}{2f\eta} \, \frac{GM}{c^2} 
  \ \approx \ 75 \, \left(\frac{0.2}{\eta}\right) 
  \, \left(\frac{0.1}{f}\right)\, \frac{GM}{c^2}.
\end{eqnarray}
We note again that $M$ depends on $f$, but not on $\alpha$.

We conclude that the fitted spectral features of ULXs 
(X-ray luminosity, temperature and ratio of thermal/non-thermal 
contribution) suggest masses $\sim 50$--$100 M_{\odot}$, at the extreme 
end of, but still consistent with models of stellar evolution.
The emitted luminosity is a few times $L_{\rm Edd}$, but 
the disk radiative contribution alone is $\la L_{\rm Edd}$. 
The rest is generated outside the disk by non-thermal processes. 
The largest contribution to the thermal disk emission comes 
from $R \approx R_{\rm c} \approx 100$ gravitational radii; 
inside this region, non-thermal processes dominate. In principle, 
this can be tested by studying the short-term variability timescale 
of the thermal and non-thermal components.
We speculate that an efficient disk-corona-outflow coupling 
with vertical transport of energy can be achieved through 
large-scale magnetic fields, when the azimuthal-vertical 
component of the magnetic stress is properly taken into 
account \cite{kun04}. See \cite{SoK06} for a more detailed 
discussion.

\begin{theacknowledgments}
RS acknowledges support from an OIF 
Marie Curie Fellowship, through University College London.
ACG acknowledges support from the 
{\it Funda\c{c}\~ao para a Ci\^encia e a Tecnologia (FCT)}, 
under grant BPD/11641/2002. 
ZK acknowledges a University of Sydney Bridging Support 
research grant.
\end{theacknowledgments}








\bibliographystyle{aipprocl} 


\end{document}